# 20 T DIPOLES AND BI-2212: THE PATH TO LHC ENERGY UPGRADE*

P.M. McIntyre, K. Damborsky, E.F. Holik, F. Lu, A.D. McInturff, N. Pogue, A. Sattarov, E. Sooby
Texas A&M University, College Station, TX 60439, U.S.A.

*Abstract*

Increasing the energy of the LHC would require a ring of ~20 T magnets using the superconductors $Nb_3Sn$ and Bi-2212/Ag. The technology for Bi-2212/Ag wire, cable, and coil has advanced significantly but is still far short of the performance needed for such magnets. New technology for both wire and cable is under development, which if successful would yield the needed performance.

## INTRODUCTION

The possibility of tripling the energy of the Large Hadron Collider was proposed in 2004 [1]. That proposal for the Tripler was motivated by its potential for new physics and by recent advances in technology that offered a path to its feasibility. An LHC Tripler would access the entire range of masses predicted for the particles of supersymmetry. The pacing technology for the Tripler was a ~24 T arc dipole. Developments at that time were encouraging: model dipoles using the low-temperature superconductor $Nb_3Sn$ [2] attained near-short-sample performance to >16 T; wire [3] and cable [4,5] using Bi-2212/Ag appeared to offer promise that 20 T might be attainable.

Today LHC is operating to produce hadron collisions at 7 TeV collision energy and is moving forward with a program to increase collision energy and luminosity to its design parameters. It is an appropriate time to revisit the potential for an LHC energy upgrade.

## BI-2212 WIRE: STALLED AT 200 A/mm$^2$

Figure 1 shows the present-day performance of conductors using NbTi, $Nb_3Sn$, Bi-2212, and YBCO, and Bi-2223. The high-temperature superconductor Bi-2212 is the only round-wire superconductor that can operate at magnetic fields beyond 18 T for dipoles. Since round wire is generally considered to be essential for a transposed high-current cable, only Bi-2212 would seem to offer the possible basis for the inner coils of a >20 T dipole. Shown are in red on Figure 1 are the working lines for the LHC dipole (using NbTi), the working line for HD1 (using $Nb_3Sn$) [2], and a working line that would be required for an LHC Tripler dipole using Bi-2212/Ag inner windings. We need an engineering current density (averaged over wire cross section) of $j_e$~600 /mm$^2$ at 24 T, 4.2 K.

In 2004 Miao [3] presented encouraging results in the development of Bi-2212/Ag round wire: $j_e$ = 400 A/mm$^2$ for 1 m long sample coils in 24 T, 4.2 K. The intrinsic performance in Bi-2212/Ag can be estimated from thin film studies, in which a layer current density of $7 \times 10^4$ A/mm$^2$ was attained [6], so there is ample room for improvement in a practical wire.

Yet six years later the state-of-art short-sample performance of Bi-2212/Ag wire is $j_e$~320 A/mm$^2$ [7], and the state-of-art performance of this same wire in coils is $j_e$~200 A/mm$^2$ in both solenoids [7] and dipoles [9]. It might seem that we are going in the wrong direction!

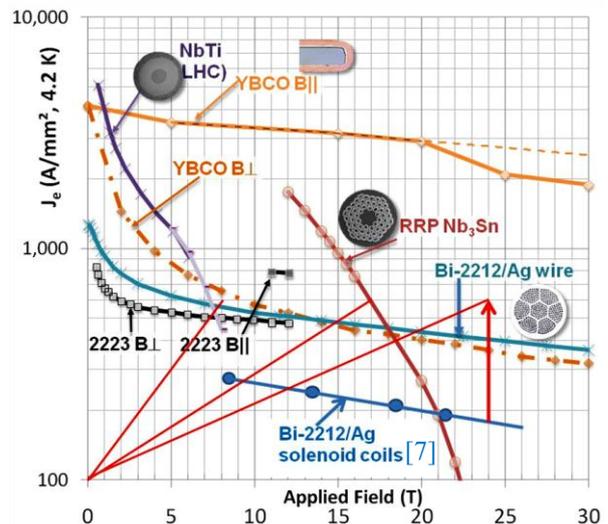

Figure 1: Recent performance of superconductors: only Bi-2212 has the potential for 20 T inner windings [8].

The earlier wire results were obtained before serious efforts had been made to make long cables and long coils from wire. Coil fabrication must be done using a wind-and-react technique, in which a high-temperature heat treatment is required to melt and re-crystallize Bi-2212 grains in the final-form coil. It was found that the earlier wire had a tendency to leak its core material during the melt phase of heat treatment so that the stoichiometry was altered and the density of the core material was depleted. These problems were helped by adding additional Ag to the wire cross-section, but this reduced the average current density accordingly. To date it has not been possible to recover the earlier $j_e$ performance. This remains a key challenge if an LHC energy upgrade is to be feasible.

## UNDERSTANDING THE LIMITS TO J$_E$

Current transport in the cores of a multi-filament Bi-2212/Ag wire is hindered by porosity and poor connectivity, both of which are largely inherent to the oxide-power-in-tube (OPIT) process used in its fabrication. In this process a fine powder of Bi-2212 is loaded and sealed into an Ag tube; the tube is drawn, restacked, and re-drawn to form the multi-filament composite shown in Figure 2. Hellstrom and co-workers [10] have studied the development of the microstructure in the cores of this wire during the heat treatment process, and from their studies a new understanding of the limitations to current transport is arising.

The wire starts life with significant porosity from the void space between randomly oriented powder particles.

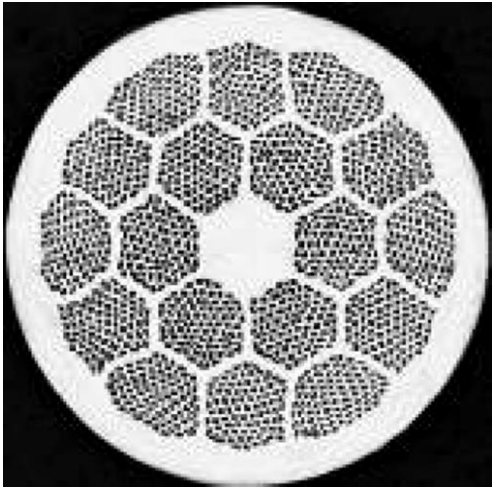

Figure 2: Cross section of OPIT Bi-2212/Ag strand [7].

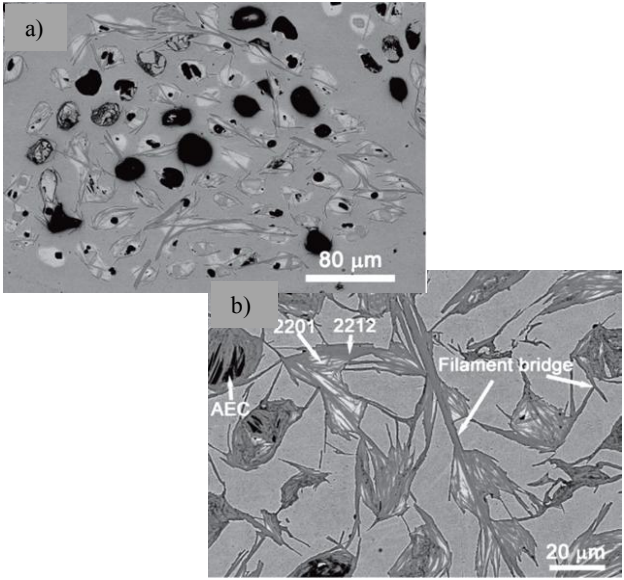

Figure 3: Scanning electron micrograph (SEM) images of core cross sections in OPIT Bi-2212/Ag wire, quenched during heat treatment: a) bubbles in melt before solidification; b) bridging, voids, and parasitic phases after annealing is complete [10].

When the powder is melted that void space becomes bubbles in the melt. The bubbles coalesce under surface tension to form large voids that span the cross section of each core. Figure 3a shows the coalesced bubbles and also the etching at Ag grain boundaries by the corrosive liquid melt. Figure 3b shows the Bi-2212 grains that have re-grown during the final anneal of the wire. Grain growth orients along the Ag interface, and the growth is fastest in the ab plane (which is the plane in which maximum current can be transported in the superconducting state). Many re-grown Bi-2212 grains now bridge between cores, and there are many high-angle grain boundaries that are problematic for supercurrent transfer from grain to grain. And so it is that the transport current density attained in OPIT wires [7] is only ~5% of that seen in thin films [6].

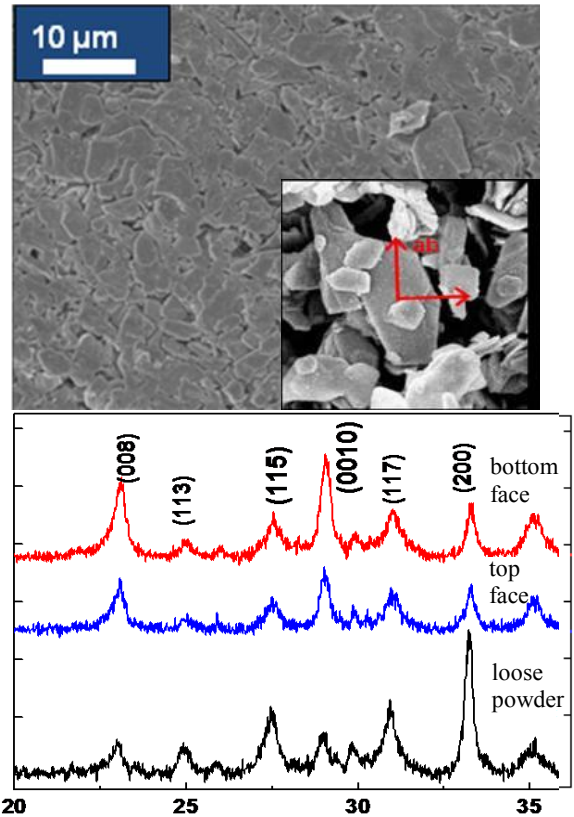

Figure 4: Study of pressed tablets of Bi-2212 fine powder: a) SEM micrograph of textured powder; b) detail showing micaceous particle morphology in loose powder; c) XRD analysis of texturing in fine powder and in pressed tablets.

## TEXTURED POWDER JELLY ROLL: HOPE FOR NEW PERFORMANCE?

The above properties of OPIT-process Bi-2212/Ag wire led us to explore an alternative process for wire fabrication that directly addresses the issues of porosity and connectivity. It begins by preparing a cold-sintered flat ribbon of Bi-2212 fine powder in which most of the grains are oriented so that their ab plane is parallel to the ribbon face. This *texturing* of the powder has two important benefits: it minimizes the porosity in the final cores of a wire, and it may eliminate the necessity to fully melt the powder during processing.

The easiest way to texture Bi-2212 powder is to press a tablet using a hydraulic press [11]. The mechanical agitation among the powder particles during compression is remarkably effective in re-arranging them into a planar texture. Figure 4a shows an SEM micrograph of the pressed powder in such a tablet. Figure 4b shows a detail of the flake-like (micaceous) Bi-2212 particles. Figure 4c shows the XRD spectra for the loose powder and for tablets pressed with 70 MPa compression. A texture parameter $\tau$ (fraction of particles aligned with ab planes parallel to tablet face) has been extracted from the XRD spectra of loose powder, the powder cores in green-state OPIT wire, and tablets pressed with various degrees of compression. The data are presented in Figure 5.

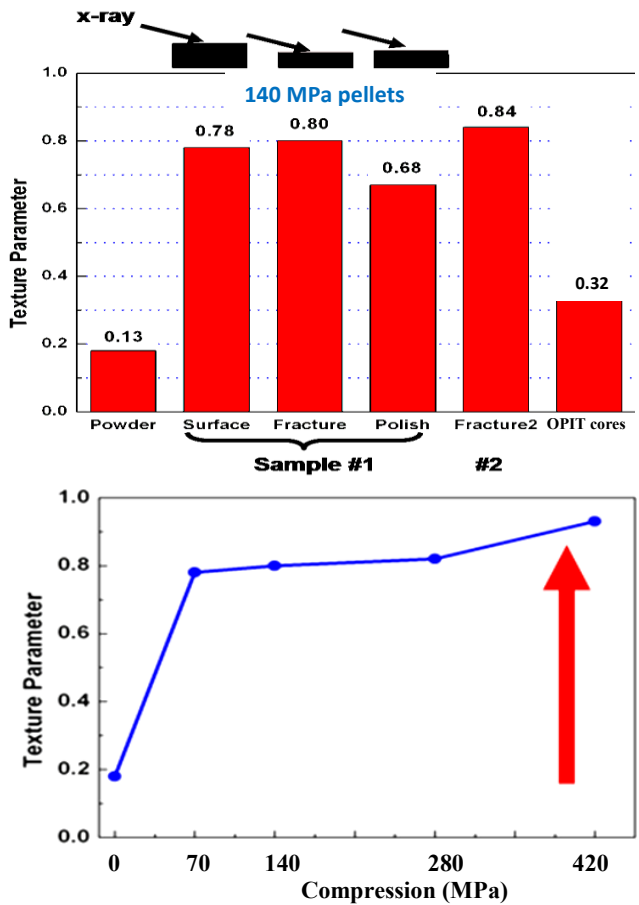

Figure 5: Texture parameter τ measured for fine-powder Bi-2212: a) loose powder, OPIT cores, and pressed tablets; b) dependence of τ on the amount of compression.

In a pellet 80% of the Bi-2212 particles are aligned with the tape face; the texture is the same in the interior of the pellet (fractured through its thickness) as on the surface; and it is largely independent of the compression (beyond 70 MPa) used to form the pellet. By contrast τ=13% for loose powder and τ=32% for the powder cores within conventional OPIT wire before heat treatment.

This development led us to conceive of an alternative method for wire fabrication in which Bi-2212 fine powder is roll-compacted to form a continuous ribbon, and the ribbon is compounded into a textured-powder 'jelly-roll' (TPJR) wire. The process begins by passing the powder though a roll-compaction system such as the Chilsonator [12], as shown in Figure 6a. The ribbons of cold-sintered Bi-2212 powder are assembled side-by-side on a ribbed Ag foil and a cover Ag foil is welded on to make a wide hermetic tape (Figure 6b). The tape is then rolled transversely, sleeved into a Cu tube, and drawn to final wire size (Figure 6c). Then the Cu is etched off to expose the final wire.

The textured micaceous powder within the laminar cores facilitates drawing of the wire. As the billet is drawn the particles in the textured powder should slide upon one another on their parallel faces and re-arrange to accommodate the area reduction.

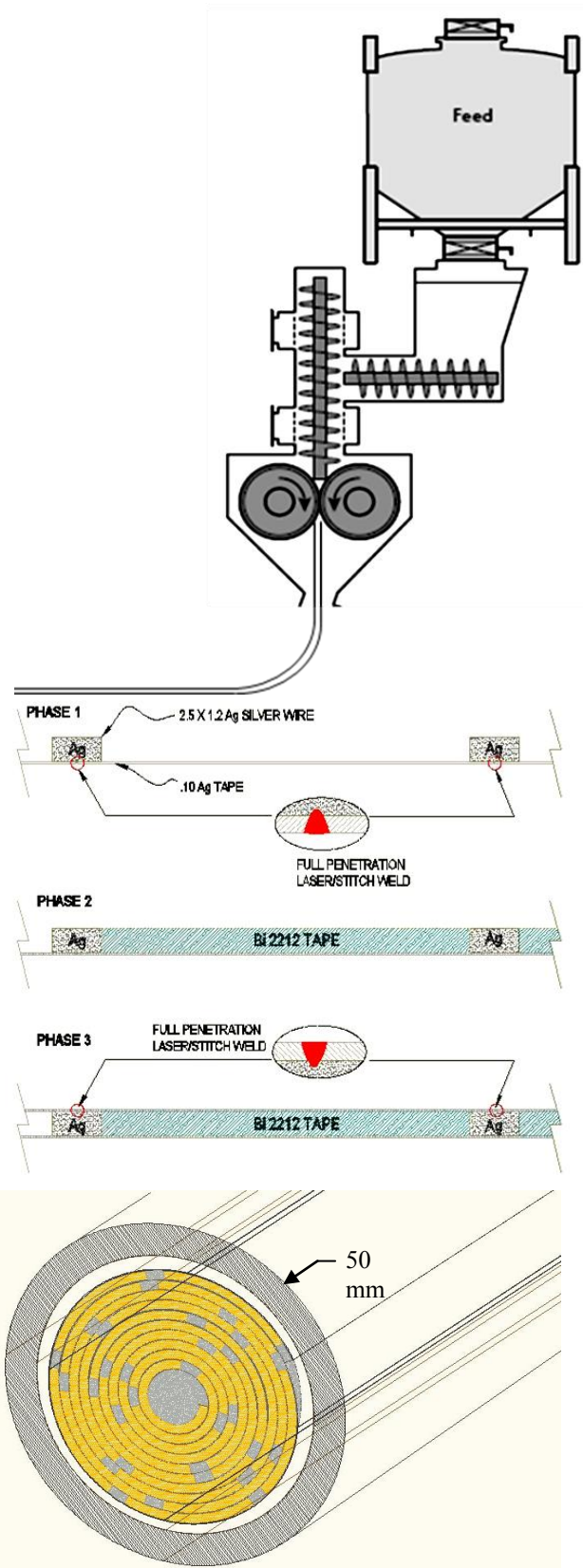

Figure 6: a) Chilsonator apparatus used to roll-compact Bi-2212 fine powder into continuous ribbons; b) incorporation of powder tape into Bi-2212/Ag tape; c) final jelly-roll round wire formed by rolling and drawing the tape.

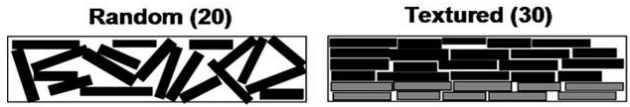

Figure 7. Illustration of the effect of texturing upon porosity in a core channel: the same channel holds 20 flakes of untextured powder, 30 flakes of textured powder.

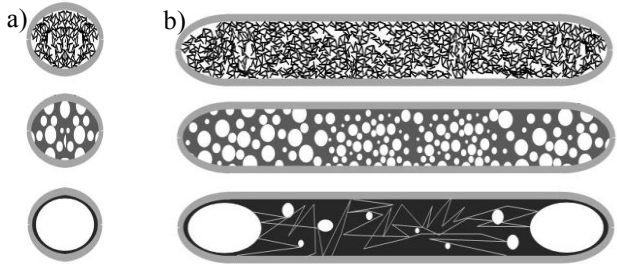

Figure 8: Illustration of bubble coalescence during full-melt heat treatment: a) full-diameter bubble forms and occludes round channel; bubbles form on the flanking edges of flat channel, leaving center free for grain growth.

We plan to evaluate possible heat treatment strategies for the TPJR wire: diffusion bonding of grains without melting, partial-melt processing, and full-melt processing. The first two methods would avoid the full-melt that is required for the OPIT process to develop texture and connectivity in its initially untextured cores. Avoiding melt would greatly reduce the issues of core leakage, cation migration, subelement bridging, and bubble coalescence that cause problems for the OPIT process. It is reasonable to hope that such non-melt treatment work well since the particles should be in face contact under compression, an ideal basis for bonding and connectivity by diffusion or partial melt.

If we find that full melt is nevertheless required, the TPJR process retains two important benefits compared with OPIT. The first benefit is improved packing. Figure 7 illustrates that there is less porosity when the flake-like particles are aligned.

The second benefit concerns the coalescence of bubbles. Figure 8a illustrates how the minimum-energy configuration of bubbles in a round channel is a large bubble that locally occludes the whole channel. Figure 8b shows the minimum-energy configuration for a highly aspected channel in the TPJR strand: bubbles coalesce on the two flanking edges of the channel, but leave the center of the channel clear for growth of textured grains of Bi-2212.

We will soon receive delivery of the roll-forming apparatus and begin development of the flat tape and jelly-roll wire. Much work is ahead to find optimum parameters for ribbon compression, tape fabrication, jelly-roll processing, and final heat treatment. The above analysis shows why we are hopeful that this approach may make possible higher $j_e$ current transport in TPJR wire.

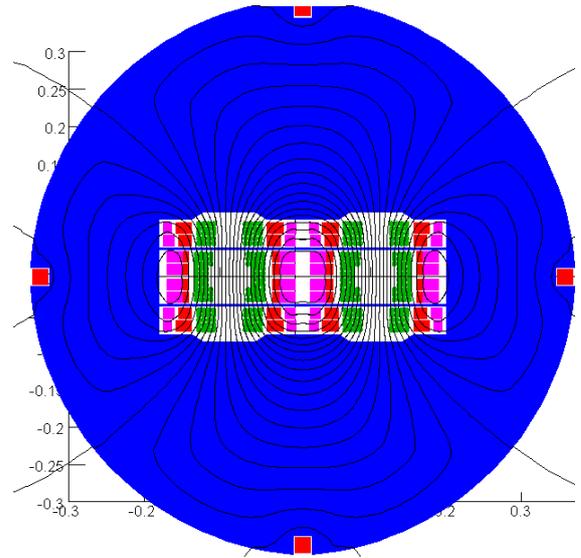

Figure 9: Cross section of dual dipole for LHC energy upgrade.

Table 1: Main Parameters of the Dual Dipole of Figure .

| | | |
|---|---|---|
| bore field (short sample) | 21 | T |
| coil current | 15 | kA |
| aperture | 50 | mm |
| stored energy/bore | 3.3 | MJ/m |
| max. stress in Nb$_3$Sn windings | 170 | MPa |
| strand cross-section/bore in coil: | | |
| Nb$_3$Sn | 52 | cm$^2$ |
| Bi-2212 | 55 | cm$^2$ |

## STRUCTURED CABLE AND 20 T DIPOLE

Figure 9 shows a conceptual design for a dual dipole that would have sufficient aperture (50 mm) for an LHC energy upgrade. The design assumes the use of Bi-2212 windings (green) in the coil region where the field strength exceeds ~16 T. It assumes that a strand performance $j_e$~800 A/mm$^2$ (20 T, 4.2 K) can be achieved in the windings, and $j_c$=2500 A/mm$^2$ (non-Cu, 12 T, 4.2K) in Nb$_3$Sn windings The Nb$_3$Sn windings are graded in wire diameter for the same $j_e(B)$ (magenta inner, red outer).

Even with the necessary current density in long-length wire and cable, it will still be necessary to protect the Bi-2212/Ag winding from strain degradation of the wires under the immense Lorentz stress produced on the windings in a 20 T dipole [13]. For this purpose we developed a structured cable [14] in which coil stress is by-passed around the fragile round wires so that no strain degradation should result. The cable is shown in Figure 10.

The Bi-2212/Ag inner winding in the coil of the dipole in Fig. 9 is a rectangular-cross-section (Fig. 11a) wound using a 16-strand structured cable (Fig. 11b) Table 1 gives the main parameters of the dual dipole. With the above assumptions of short-sample wire performance, the short-sample limit of the dipole is 21 T.

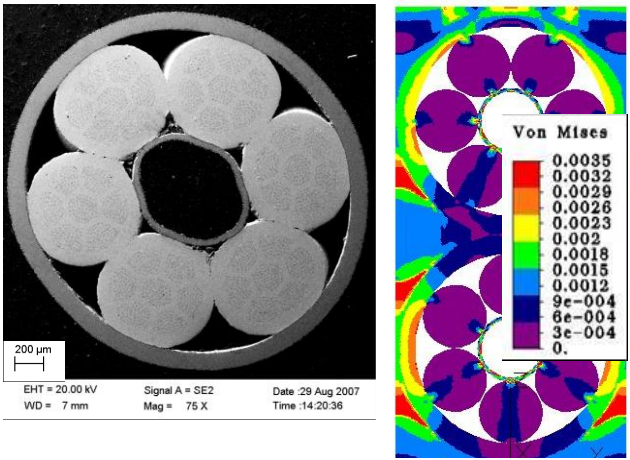

Figure 10: Structured cable of Bi-2212 round strands: a) micrograph of cross section; b) von Mises strain in structured cable when 100 MPa external load is applied [5].

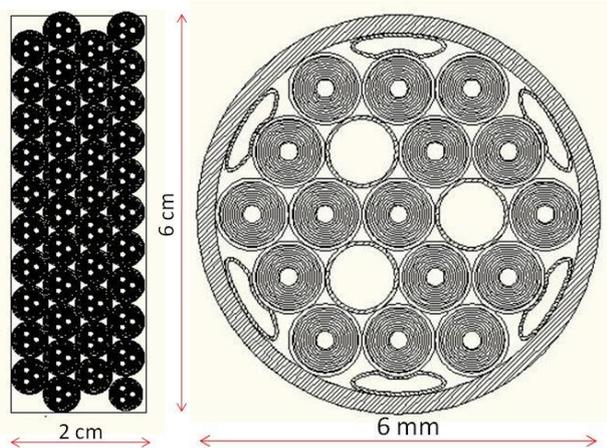

Figure 11: Cross-section of Bi-2212 inner winding, wound with 16-strand structured cable using TPJR round strands.

## CONCLUSIONS

In the examination of the many challenging elements of an LHC energy upgrade that were presented at this EuCARD workshop, the high-field arc dipoles appear to present the biggest challenge at present. The critical technology for these dipoles is the current density in Bi-2212/Ag round wire and the degradation of that wire when made into thick windings and loaded with high Lorentz stress.

An alternative method for Bi-2212/Ag wire fabrication is described, which holds the potential for enabling a further improvement in current density. A design for structured cable is presented that manages stress within the coils of the Bi-2212/Ag inner windings so that it cannot accumulate to levels that would degrade performance.

The viability of an LHC energy upgrade will depend upon the success of these developments and similar ones by other authors. It is to be hoped that a successful outcome can be matured in time for consideration of an energy upgrade of LHC after its first decade of high-luminosity physics running.

This work is supported in part by the US Dept. of Energy, grant DE-FG03-95ER40924, and by an endowment from the Cynthia and George Mitchell Family Foundation.